# Robust topological interface states in a lateral magnetic-topological heterostructure

*Qun Niu, Jie Yao, Quanchao Song, Humaira Akber, Qin Zhou, Xiaofang Zhai, and Aidi Zhao\**

Qun Niu, Quanchao Song, Qin Zhou, Xiaofang Zhai, and Aidi Zhao
School of Physical Science and Technology, ShanghaiTech University, Shanghai 201210, China
zhaoad@shanghaitech.edu.cn

Jie Yao, Humaira Akber
Department of Chemical Physics, University of Science and Technology of China, Hefei, Anhui 230026, China

**Abstract:** Introducing uniform magnetic order in two-dimensional topological insulators (2D TIs) by constructing heterostructures of TI and magnet is a promising way to realize the high-temperature Quantum Anomalous Hall effect. However, the topological properties of 2D materials are susceptible to several factors that make them difficult to maintain, and whether topological interfacial states (TISs) can exist at magnetic-topological heterostructure interfaces is largely unknown. Here, we experimentally show that TISs in a lateral heterostructure of $CrTe_2$/Bi(110) are robust against disorder, defects, high magnetic fields (time-reversal symmetry breaking perturbations), and elevated temperature (77 K). The lateral heterostructure is realized by lateral epitaxial growth of bilayer (BL) Bi to monolayer $CrTe_2$ grown on HOPG. Scanning Tunneling Microscopy and non-contact Atomic Force Microscopy demonstrate a black phosphorus-like structure with low atomic buckling (less than 0.1 Å) of the BL Bi(110), indicating the presence of its topological properties. Scanning tunneling spectroscopy and energy-dependent d$I$/d$V$ mapping further confirm the existence of topologically induced one-dimensional in-gap states localized at the interface. These results demonstrate the robustness of TISs in lateral magnetic-topological heterostructures, which is competitive with those in vertically stacked magnetic-topological heterostructures, and provides a promising route for constructing planar high-density non-dissipative devices using TISs.

# 1. Introduction

Heterostructures of topological insulators (TIs) and magnetic materials are highly intriguing for realizing exotic topological quantum phases, including the quantum anomalous Hall (QAH) effect. QAH effect is determined by both the magnetic order and topological properties of materials, where the non-dissipative chiral conducting states propagate along the edges, holding potential applications in ultra-low-power electronic devices.[1–4] Initially achieved in magnetically doped TIs, the QAH effect was limited to a low realization temperature due to the disorder introduced by magnetic dopants.[5–10] Recently, the QAH effect has also been observed in three-dimensional (3D) magnetic topological material $MnBi_2Te_4$,[11,12] and Moiré heterostructures.[13,14] Beyond these systems, a more favorable option is to construct TI-magnet heterostructures, introducing uniform magnetic order into TIs via magnetic proximity effects to achieve higher-temperature QAH effects.[15–18] This strategy has been implemented in 3D TI-ferromagnet/antiferromagnet heterostructures,[19–21] resulting in a notably enhanced gap. Compared to 3D TIs, 2D TIs feature tunable electronic properties,[22–26] larger topological gaps,[27–30] and nanometer-sized one-dimensional (1D) conductive channels, promising high-temperature QAH effect and miniaturized electronic devices. Furthermore, the 2D nature of these materials allows for designing heterostructures through lateral and vertical stacking, providing rich tunability and flexibility for constructing practical devices. Among the 2D materials, few layers of its two stable allotropes, Bi(111) and Bi(110),[31–33] have both been demonstrated to be large-gap 2D TIs owing to the strong spin-orbit coupling (SOC).[26,30,34–37] BL Bi(110) films on HOPG have been demonstrated to be topologically non-trivial with a large bulk gap and significant in-gap states localized at the edges, *i.e.*, topologically induced edge states (TESs).[26] However, the TES has been reported to be suppressed by edge reconstruction, in-plane ferroelectricity, strong interlayer coupling, or other factors.[22,25,35,36,38] Whether the TES survives at the interface and how it evolves in lateral or vertical heterostructures consisting of BL Bi(110) and 2D magnetic materials are unknown.

In this work, we address this issue by constructing both lateral and vertical heterostructures of BL Bi(110) and monolayer (ML) $CrTe_2$. ML $CrTe_2$ is a typical metallic 2D van der Waals (vdWs) magnet.[39–43] For lateral heterostructure of BL Bi(110)/ML $CrTe_2$, high-resolution scanning tunneling microscopy (STM) revealed that the interfaces between $CrTe_2$ and Bi(110) are seamless and atomically flat, devoid of reconstruction. Atomic-resolved non-contact atomic force microscopy (NC-AFM) disclosed that the height difference between the two neighboring surface

Bi atoms is less than 0.1 Å. Such a low-buckled atomic structure is in line with the theoretically predicted topologically non-trivial nature of Bi(110). Scanning tunneling spectra (STS) demonstrated pronounced in-gap states at the interfaces as well as an insulating character of the BL Bi(110) in bulk. The real-space distribution and robustness of the in-gap states were further verified by visualizing the continuous 1D conducting channels at the interfaces with d$I$/d$V$ mappings at 77 K and 5 K with the external magnetic field up to 4T, irrespective of local disorder or defects. In contrast, no chiral/helical TESs of Bi(110) were found in vertical heterostructures due to strong interlayer coupling interactions with $CrTe_2$. Our research provides a new strategy to introduce topological-induced interfacial states (TISs) of BL Bi(110) by constructing a lateral heterostructure free of reconstruction. The TISs of Bi(110) in the lateral heterostructure are experimentally demonstrated to be robust against disorder, defects, high magnetic fields (time-reversal symmetry breaking perturbations), and elevated temperature (77 K), providing a promising playground for constructing planar high-density non-dissipative devices.

## 1. Results

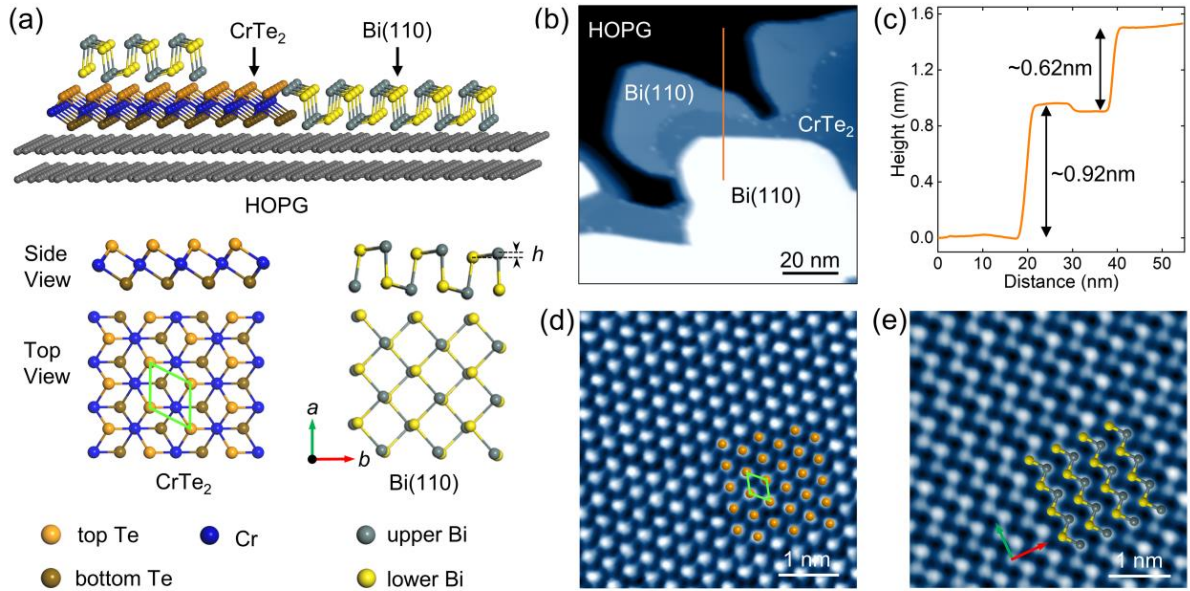

**Figure 1.** Fabrication of BL Bi(110) and $CrTe_2$ heterostructures by a two-step molecular beam epitaxy method. (a) Schematic view of $CrTe_2$, Bi(110), and heterostructures. The unit cell of $CrTe_2$ is marked with a bright green diamond. In Bi(110), the dark green arrow points towards the short axis-***a***, and the red arrow points towards the long axis-***b***. (b) STM topographic image of the heterostructures grown on HOPG ($V_b = -1.6$ V, $I_t = 50$ pA). (c) Line profile taken along the orange
3

line in (b). (d) Atomic resolution of ML CrTe$_2$ grown on HOPG ($V_b$ = 0.4 V, $I_t$ = 200 pA), showing a hexagonal lattice. (e) Atomic resolution of BL Bi(110) grown on HOPG ($V_b$ = −10 mV, $I_t$ = 800 pA), showing a BP-like lattice.

Bi(110) and CrTe$_2$ heterostructures with low atomic buckling structure. Ultrathin Bi(110) and CrTe$_2$ lateral and vertical heterostructures were fabricated on HOPG via molecular beam epitaxy (MBE), with the structural schematics depicted in **Figure 1a**. CrTe$_2$ features a vdWs layered structure. Each layer of CrTe$_2$ constitutes chromium cations centering at tellurium octahedrons (as illustrated in the lower left panel in Figure 1a), belonging to the space group *P3m1*.[39,44] On the contrary, Bi(110) exhibits a black phosphorus (BP)-like structure (shown in the lower right panel in Figure 1a). Each surface unit cell of Bi(110) contains two atoms positioned at the corner and near the center of the rectangular unit cell, respectively.[26] Typically, there is a height difference *h* between the two atoms, and this value is critical to the topological and ferroelectric properties.[22,26,45] Figure 1b presents a large-scale STM image of the grown ultra-thin film, where Bi(110), CrTe$_2$, and the lateral heterostructure interface are clearly identifiable. The height of the BL Bi(110) and ML CrTe$_2$ films grown on HOPG is approximately 0.92 nm and 0.85 nm, separately (Figure 1c), which is higher than the theoretical height of 0.655 nm[31] and 0.597 nm,[46] respectively, suggesting a large vdWs gap with HOPG substrate. BL Bi(110) deposited on CrTe$_2$ exhibits an apparent height of 0.62 nm, consistent with theory[31] (Figure 1c). The atomic-resolved STM image of ML CrTe$_2$ displays a hexagonal lattice (Figure 1d), with an in-plane constant measured as *a* = 0.362 nm, as reported in our previous work.[40] Figure 1e shows the atomic resolution of BL Bi(110) on HOPG. The in-plane lattice constants are *a* = 0.452 nm and *b* = 0.473 nm, respectively, which closely match those of freestanding BL Bi(110) (*a* = 0.454 nm and *b* = 0.475 nm),[31] indicating the weak interaction between Bi(110) films and HOPG, and the negligible lattice distortion. In each surface unit cell of the Bi(110), both Bi atoms can be clearly identified (Figure 1e), and we emphasize that the relative height difference between the two atoms is pivotal for topological and ferroelectric properties,[22,26,45] hence it is essential to acquire the accurate height information between different Bi atoms.

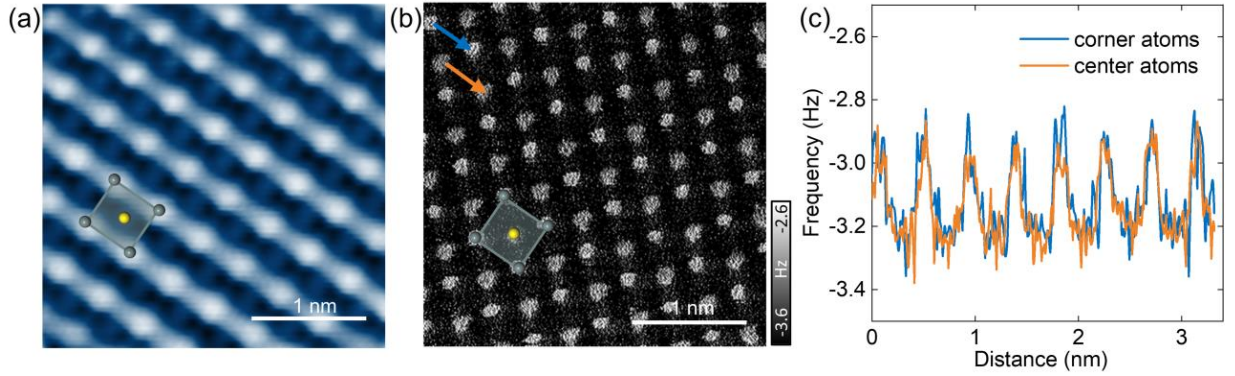

**Figure 2.** BL Bi(110) film with a low-buckled BP-like structure. (a) High-resolution STM image of BL Bi(110) ($V_b$ = 0.3 V, $I_t$ = 1.7 nA). The surface unit cell is marked with grayish-blue rectangle, and the corner atoms and center atoms are marked with grayish-blue and yellow spheres, respectively. (b) Corresponding NC-AFM image in constant height mode. (c) Line profiles obtained along the corner Bi atoms (blue arrow in (b)) and central Bi atoms (orange arrow in (b)).

However, precise measurement of atomic height difference directly from STM images is not feasible due to the contribution from the local density of states (LDOS) to the apparent atomic height. To overcome this problem, NC-AFM with a qPlus sensor was employed to image the surface of Bi(110) (**Figure 2**). Figures 2a and 2b depict a constant-current STM image and a constant-height NC-AFM image acquired at the same area of BL Bi(110) using the same tip, with Bi atoms marked on both images with grayish-blue or yellow spheres. Figure 2c shows line profiles collected along the center and corner atoms (indicated by the orange and blue arrows in Figure 2b). The two Bi atoms within the surface unit cell exhibit nearly identical frequency shifts ($\Delta f$) (see Figures 2b and 2c). In NC-AFM, $\Delta f$ reflects the interaction force ($F$) and distance ($d$) between the tip and sample atoms.[47–49] The nearly identical $\Delta f$ values for corner and center Bi atoms demonstrate very close tip-sample atom distances, indicating low atomic buckling ($h$). In ferroelectric Bi(110) films, the $\Delta f$ difference between corner and center Bi atoms exceeds 1.4 Hz and $h$ = 0.4 Å.[22] In our sample, the $\Delta f$ difference is less than 0.06 Hz, suggesting that the $h \ll 0.4$ Å in our Bi(110) films. Since the elements at different lattice positions are the same, the force curves $F(d)$ or $\Delta f(d)$ obtained from different Bi atoms should be identical.[22,50] This allows us to estimate the atomic buckling $h$ from the same $\Delta f(d)$ curve. Putting the $\Delta f$ values extracted from Figure 2c back into the $\Delta f(d)$ curve, we estimate $h < 0.1$ Å (Figure S1). Such a low buckling strongly indicates that the BL Bi(110) in the lateral heterostructure is a 2D TI with a large bulk gap and pronounced TES according to previous work which pointed out a critical value of 0.1 Å for the nontrivial 2D TI phases by both theoretical calculations and experiments.[26]

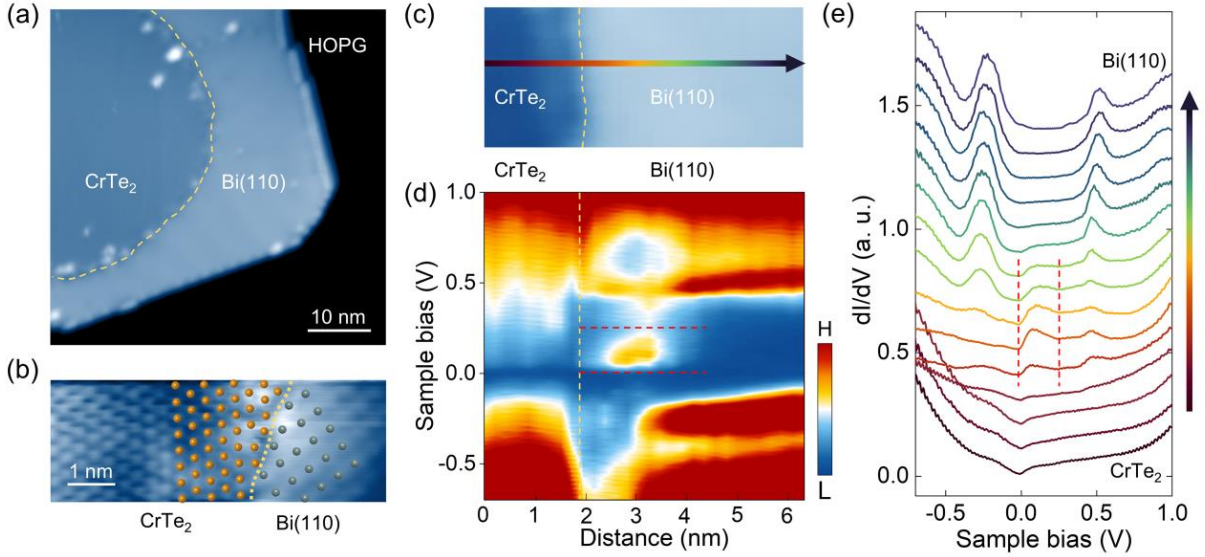

**Figure 3.** Topological interfacial states exist at the interfaces of lateral heterostructures without atomic reconstruction. (a) STM topographic image of ultrathin $CrTe_2$/BL Bi(110) lateral heterostructure film ($V_b$ = -1.6 V, $I_t$ = 50 pA). The lateral interface is marked with a yellow dashed line. (b) Schematic and high-resolution STM image of $CrTe_2$/BL Bi(110) lateral heterostructure ($V_b$ = 1.0 V, $I_t$ = 250 pA). Grayish-blue and orange spheres represent Bi and Te atoms, respectively. Only the topmost atoms are labeled. (c) STM image of lateral heterostructure ($V_b$ = 100 mV, $I_t$ = 200 pA). (d) 2D plot of the line-cut $dI/dV$ spectra taken across the lateral heterostructure interface, along the colored dashed arrow in (c) ($V_b$ = 1.0 V, $I_t$ = 200 pA). The red dashed lines mark the in-gap states of Bi(110). (e) A series of selective subset $dI/dV$ spectra corresponding to the colored arrow in (c) ($V_b$ = 1.0 V, $I_t$ = 200 pA).

To further verify the nontrivial topological properties of the BL Bi(110) in the heterostructure, we perform STM/STS measurement across the heterojunction of ML $CrTe_2$ and BL Bi(110) to confirm the existence of TISs at the lateral interfaces. **Figure 3** illustrates the atomic-scale structure and electronic properties of the $CrTe_2$/Bi(110) lateral heterostructure. A large-scale STM image of lateral heterostructure is shown in Figure 3a. The ML $CrTe_2$ island is surrounded by BL Bi(110) film and the interface is highlighted by a yellow dashed line. A high-resolution STM image shows the typical atomic structure of the lateral heterostructure interface with marks of topmost Bi atoms and Te atoms superimposed on it, as shown in Figure 3b. Interestingly, although the lattice symmetries of the two materials are distinct and the relative crystallographic orientation between them is random (see Figure S2), the interface between them is atomically seamless and without forming obvious reconstruction and dislocation on the Bi(110) side. Such a continuous epitaxy of Bi(110) is expected owing to the highly flexible Te-Bi chemical bonding between terminal Te atoms of $CrTe_2$ and Bi atoms of BL Bi(110). We further recorded a set of STS spectra across the interface from $CrTe_2$ to Bi(110) along the colored arrow in Figure 3c. Figures 3d and 3e depict a 2D plot and a series of selected tunneling spectra obtained along the colored arrow, respectively. ML $CrTe_2$ exhibits typical metallic behavior with finite LDOS near the Fermi level, in agreement with previous results.[39,40] The d$I$/d$V$ spectra taken on BL Bi(110) far away from the interface exhibit a typical gapped semiconducting behavior with two pronounced peaks at around −0.2 V and 0.5 V, corresponding to the valence and the conduction band edges.[26],[38] Remarkably, significantly enhanced DOS rises in the band gap in the spectra taken on the Bi(110) side of the interface (highlighted by red dashed lines in Figures 3d and 3e). These in-gap states localized in the same energy range (0 V~ +0.25 V) as the in-gap states previously identified as non-trivial TES in BL Bi(110) on HOPG,[26] suggesting the same topological origin.

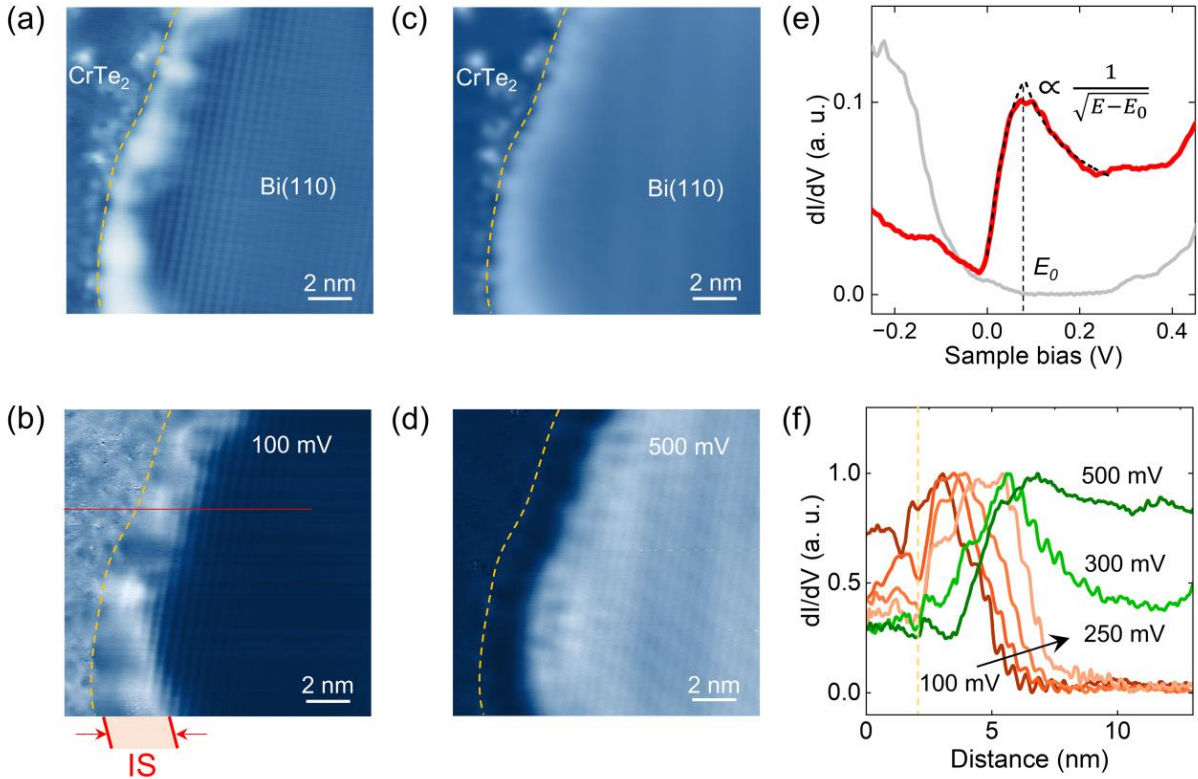

**Figure 4.** The robustness and 1D nature of topological interfacial states in the lateral

heterostructure. (a), (b) STM image and corresponding STS mapping of CrTe$_2$/BL Bi(110) lateral heterostructure recorded at $V_b$ = 100 mV (within the bulk band gap) ($I_t$ = 200 pA). The yellow dashed lines mark the lateral interface position. (c), (d) STM image and corresponding STS mapping of CrTe$_2$/BL Bi(110) lateral heterostructure recorded at $V_b$ = 500 mV (outside the bulk band gap) ($I_t$ = 200 pA). (e) 1D LDOS fitting (black dashed line) of the measured interface spectrum (red solid line) ($V_b$ = 1.0 V, $I_t$ = 200 pA). (f) LDOS profiles measured at different energies along the red line (marked in (b)) perpendicular to the lateral interface.

There are several possibilities for the enhanced in-gap DOS at interfaces with trivial origin from local defects/dislocations, or unsaturated dangling bonds. In order to distinguish the non-trivial TISs from these trivial interface states, d$I$/d$V$ mappings were performed at the CrTe$_2$/Bi(110) interface (**Figure 4**). The TISs should be essentially structure-irrelevant and continuously spread in real space with a typical 1D character, while those trivial interface states are strongly structure-relevant and localized at local defects/dislocations with non-uniform atomic structure. Figures 4a-d and Figure S3 show typical STM images and corresponding d$I$/d$V$ maps recorded at different sample biases for a lateral heterostructure, with the interface highlighted by a yellow dashed line. At sample biases $V_b$ of 100 to 250 meV, the in-gap interface states do manifest as a continuously spread 1D conducting channel along the interface on the Bi(110) side, seen as the enhanced contrast areas in both STM images and d$I$/d$V$ maps (Figures 4a, 4b and Figures S3 a-d and g-j). The width of the spatial distribution of the interface state (~3 nm) is consistent with the non-trivial origin demonstrated previously[26] and observed in other systems.[51–53] At sample biases higher than 250 meV, the interface states eventually retract from the interface and merge into the bulk states which rise at $V_b$ > 300 meV and spatially disperse in the area ~ 2 nm away from the interface (Figure 4d). Moreover, the in-gap states in the d$I$/d$V$ spectrum taken at the interface can be perfectly fitted by the 1D parabolic LDOS formula $D \propto 1/\sqrt{E - E_0}$, confirming their 1D nature (as shown by the black dashed line in Figure 4e).[22,38] As pointed out by recent literature, the topological nature of the 1D states necessarily dictates a specific spatial and energy dependence of the TISs.[54] In particular, with increasing energy, the localization width of TISs must continuously increase from lower energy to higher energy reaching the gap edge. To visualize this evolution, in Figure 4e, we plot spatial profiles measured at different energies along the red line perpendicular to the interface from the d$I$/d$V$ maps (Figure 4b and Figure S3) taken at a set of energies from 100 mV up to 500 mV. The energy-dependent delocalization behavior (indicated by the black arrow) essentially conforms to the structure-irrelevant nature of the observed in-gap interface states. Combining the 1D characteristic of the spectra shape, the continuous spatial distribution of the in-gap states localized along the interface, as well as the energy-dependent evolution of the spatial distribution, we conclude that the observed in-gap states are truly topologically-induced non-trivial 1D interface states.

Intrinsically, the topologically-induced interface states consist of two counterpropagating helical edge channels (spin-up and spin-down) in real space which are protected by time-reversal symmetry. When a magnetic field is applied perpendicular to the 2D plane, the in-gap states measured at the edge are expected to be substantially suppressed and a gap at the edge-state Dirac point is expected to be opened. [55,56] To verify the effect of external magnetic perturbation, we investigated the TISs under a magnetic field up to 4 T. As displayed in Figure S4, STS mappings and individual spectra were collected under a magnetic field of 4 T applied perpendicular to the sample surface, and no significant effect was observed on these in-gap states, demonstrating the robustness of TISs against external magnetic fields. Such robustness is consistent with a single chiral channel (only one spin orientation) in the presence of an external magnetic field,[3,10,11,20] which can be essentially regarded as a 2D version of quantum anomalous Hall edge states. Considering the $CrTe_2$ is a typical magnetic material, it is highly possible that the TISs on the edge of Bi(110) are such chiral states since the penetration length of the magnetic proximity effect is expected to be limited to a few nanometers which exactly encompasses the measured width of 3 nm for the TISs channel.

The large bulk gap of Bi(110) and the high curie temperature of $CrTe_2$ (> 77 K[41,57]) should in principle protect the topologically-induced interface channel up to 77 K. In our experiments, we indeed observe the preservation of the TISs in STS measured at 77 K, as shown in Figure S5. Such robustness against elevated temperatures highlights the potential of the TISs for low-cost applications.

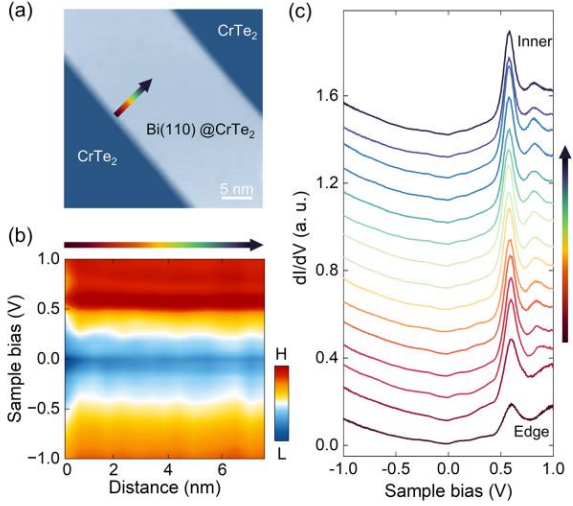

**Figure 5.** Morphology and electronic properties of Bi(110)@CrTe$_2$ vertical heterostructures. (a) STM image of Bi(110) deposits on CrTe$_2$ ($V_b$ = -1.3 V, $I_t$ = 30 pA). (b), (c) Typical d$I$/d$V$ spectra of Bi(110) in vertical heterostructure ($V_b$ = -1.3 V, $I_t$ = 30 pA). 2D plots (b) and individual (c) of line-cut d$I$/d$V$ spectra taken perpendicular to the edge along the colored arrow labeled in (a).

The conservation of the nontrivial TISs benefited from the minimized coupling between the electronic states of CrTe$_2$ and Bi(110) in the lateral geometry. In contrast, the electronic properties of BL Bi(110) are strongly altered by interlayer coupling in the vertical heterostructure of Bi(110)@CrTe$_2$, as shown in **Figure 5**. A series of STS spectra taken on the BL Bi(110) on CrTe$_2$ along the colored arrows in Figure 5a are plotted in Figures 5b and 5c. The spectra taken at the edge are nearly identical to those taken at the inner part of Bi(110) and show no edge state features (enhanced LDOS) near the Fermi level. Such a global metallic behavior is distinct from the topological insulating behavior of Bi(110) in the lateral heterojunctions. It is not surprising that numerous previous studies have shown that the topological properties of 2D materials are susceptible to various factors.[25,35,38,58] Substrate or interlayer interactions may substantially modulate the electronic properties of layered materials, and strong interlayer coupling could even suppress TES completely. For example, the superlattice modulation of the TES in BL Bi(111)@Bi(110) vertical heterostructures and the edge states suppression caused by strong interface coupling in BL Bi(111)@bulk Bi.[25,35] In the BL Bi(110)@CrTe$_2$ vertical heterostructures, the electronic properties of Bi(110) are inevitably altered by interlayer coupling interaction. Such strong interaction can even be evidenced by the absence of edge reconstruction, as shown in Figure S6, in contrast to BL Bi(110) on HOPG. The breakdown of TES in the vertical heterostructures further demonstrates the advantage of lateral heterostructures in stabilizing and preserving the topological properties of layered materials.

## 3. Discussion

As previously reported, freestanding Bi(110) films grown on many substrates inevitably experience atomic reconstruction at the edges,[38,59,60] modifying the LDOS at the edges and complicating the identification of TES. Reconstructed edges are also present in our $CrTe_2$/Bi(110) lateral heterostructures in addition to interfaces without atomic reconstruction. To confirm the stabilizing effect of $CrTe_2$ on the topological properties of BL Bi(110) and the origin of edge states at the reconstructed edges, we collected STS data at the reconstructed edges and compared it with that obtained at the lateral interfaces. Figure S7a shows a typical STM image obtained at the reconstructed edge. The atomic reconstruction with a period of 4×1 can be easily observed at the edge, where part of the line profile of the reconstructed edge is presented in Figure S7b. Figures S7c-S7e give d$I$/d$V$ spectra obtained along the colored arrow (as labeled in Figure S7a) perpendicular to the reconstructed edge. Slightly enhanced in-gap LDOS also can be detected at the reconstructed edge whose intensity is much lower than the TISs at the interface, as indicated by the red dashed lines. The origin of the in-gap states at the free edge remains controversial: It is possible that the topologically-induced edge states are significantly suppressed by the reconstruction, or they could also possibly originate from reconstruction-modified trivial edge states.[38] Figure S8 depicts the STS mappings obtained at the free edge at different sample biases. These in-gap edge states show strong modulation with a periodicity matching that of the atomic reconstruction, suggesting a structure-dependent trivial origin or contribution.

## 4. Conclusion

In summary, we have successfully achieved high-quality lateral heterostructures made of BL Bi(110) and ML $CrTe_2$, and confirmed the robustness of TISs at the heterostructure interface against disorder, defects, external magnetic fields, and elevated temperatures (77 K). With the help of STM, the atomic-flat and non-reconstructed $CrTe_2$ and Bi(110) heterostructure interfaces were identified. NC-AFM revealed that the height difference between the two Bi atoms in the surface unit cell is less than 0.1 Å, structurally indicating the topological non-trivial properties of Bi(110) in the heterostructure. STS revealed the bulk semiconductor properties of Bi(110) thin films and continuous 1D conductive channels at the interface, confirming the real-space 1D distribution of the in-gap states regardless of local disorder or defects. STS obtained under a 4T magnetic field and at 77 K further demonstrated the robustness of TISs against TRS-breaking perturbations and elevated temperatures. The lateral heterostructures effectively stabilize the TISs at the interfaces by preventing the reconstruction and distortion of Bi atoms as compared to the reconstructed edge. As a comparison, the BL Bi(110) on $CrTe_2$ in the vertical heterostructure shows a trivial metallic behavior, without any topologically-induced edge properties. Our study provides the first evidence for the robustness of the TISs of Bi(110) in TI-magnet lateral heterostructures and provides new insights for constructing planar high-density non-dissipative devices using the TISs of 2D materials as well as their lateral heterostructures.

## 5. Experimental Section

Preparation of samples: HOPG (SPI-1 grade, SPI Supplies) is used as the substrate. It is cleaved in the atmospheric environment and degassed at 400 °C for 3 hours in the homebuilt MBE ultra-high vacuum chamber (base pressure 3.0 ×$10^{-10}$ mbar). $CrTe_2$ thin films were prepared by co-evaporating Cr (Alfa Aesar, 99.996%) and Te (Alfa Aesar, 99.999%) powders from two effusion cells (Beijing Quantech). During the deposition, the HOPG substrate was kept at 300 °C and the Cr source and Te source were maintained at 1000 °C and 300 °C, respectively. After depositing $CrTe_2$, the sample is annealed at 300 °C for 20 minutes, and then the substrate is decreased to room temperature to continue depositing Bi(110) films. The evaporation temperature of Bi powders (Alfa Aesar, 99.999%) is 460 °C. Then the heterostructures can be obtained.

STM/S measurements: The measurements were carried out in an ultra-high vacuum (UHV) low-temperature STM (Polar, Scienta Omicron GmbH), equipped with ± 5 T magnetic field perpendicular to the sample surface (base pressure 1.0 × $10^{-10}$ mbar). The STM and NC-AFM tips are prepared using chemically etched W-wires and cleaned in vacuum. The resonant frequency of the Qplus sensor used is $f_0$ = 25,404 Hz, with a sensor stiffness of $k_0$ ≈ 1,800 N/m and a quality factor $Q$ = 198,338. The tunneling spectra were collected through a lock-in amplifier with a sinusoidal modulation of 12 mV and 759 Hz. All data was obtained at 4.5 K or 77 K.

**Supporting Information**

Supporting Information is available from the Wiley Online Library or from the author.

**Acknowledgements**

This work was supported by the Startup Fund of ShanghaiTech University.

**Conflict of Interest**

The authors declare no conflicts of interest.

**Table of Contents**

Ultrathin CrTe$_2$/Bi(110) lateral heterostructure is epitaxially prepared on HOPG surfaces. Using scanning tunneling microscopy we demonstrate that the topological interface states in the lateral heterostructure are robust to disorder, defects, high magnetic fields (time-reversal symmetry-breaking perturbations) and high temperatures (77 K), providing a promising playground for constructing planar high-density non-dissipative devices.

*Qun Niu, Jie Yao, Quanchao Song, Humaira Akber, Qin Zhou, Xiaofang Zhai, and Aidi Zhao\**

**Robust topological interface states in a lateral magnetic-topological heterostructure**

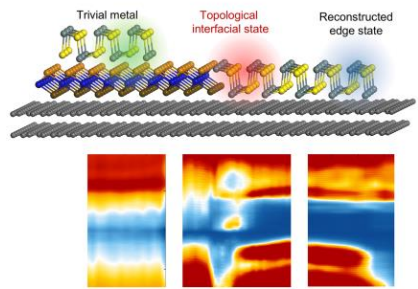

Table of contents graphic.